\documentclass[]{aa}
\usepackage[utf8]{inputenc}
\usepackage{txfonts}
\usepackage{graphicx}
\usepackage{booktabs}
\usepackage[T1]{fontenc}
\usepackage{url}
\usepackage[caption=false]{subfig}
\usepackage{float}
\usepackage{natbib}
\bibpunct{(}{)}{;}{a}{}{,}
\usepackage[colorlinks=true, citecolor=blue, linkcolor=blue, urlcolor=blue, anchorcolor=blue]{hyperref}
\newcommand{\inl}[1]{\textsf{{#1}}}
\usepackage{ulem}
\renewcommand{\arcsec}{\ensuremath{^{\prime\prime}}}
\begin{document}
\title{Mining the Kilo-Degree Survey for solar system objects}
  \author{M. Mahlke\inst{1}
    \and
        H. Bouy\inst{2, 3}
    \and
        B. Altieri\inst{1}
    \and
        G. Verdoes Kleijn\inst{4}
    \and
        B. Carry\inst{5, 6}
    \and
        E. Bertin\inst{7}
    \and \\
        J. T. A. de Jong\inst{8}
    \and
        K. Kuijken\inst{8}
    \and
        J. McFarland\inst{4}
    \and
        E. Valentijn\inst{4}
          }

  \institute{European Space Astronomy Centre (ESA/ESAC), Camino Bajo del Castillo s/n, E-28692 Villanueva de la Ca\~nada, Madrid, Spain
              \email{max.mahlke@rwth-aachen.de}
        \and
        Laboratoire d’astrophysique de Bordeaux, Univ. Bordeaux, CNRS, B18N, Allée Geoffroy Saint-Hilaire, 33615 Pessac, France
        \and
             Centro de Astrobiologia (INTA-CSIC), ESAC, Camino Bajo del Castillo s/n, E-28692 Villanueva de la Ca\~nada, Madrid, Spain
        \and
        Kapteyn Astronomical Institute, University of Groningen, Postbus 800, 9700 AV, Groningen, The Netherlands
        \and
        Universit\'e C{\^o}te d'Azur, Observatoire de la C{\^o}te d'Azur, CNRS, Lagrange, France
        \and
        IMCCE, Observatoire de Paris, PSL Research University, CNRS, Sorbonne Universit{\'e}s, UPMC Univ Paris 06, Univ. Lille, France
        \and
        Institut d'Astrophysique de Paris, CNRS UMR 7095 and UPMC, 98bis bd Arago, F-75014 Paris, France
        \and
        Leiden Observatory, Leiden University, P.O.Box 9513, 2300 RA Leiden, The Netherlands}

  \date{Received 03/04/2017; accepted 24/10/2017}

\abstract{The search for minor bodies in the solar system promises insights into its formation history.  Wide imaging surveys offer the opportunity to serendipitously discover and identify these traces of planetary formation and evolution. 
}{We aim to present a method to acquire position, photometry, and proper motion measurements of solar system objects in surveys using dithered image sequences. The application of this method on the Kilo-Degree Survey is demonstrated.
}{Optical images of 346\,${\rm deg}^2$ fields of the sky are searched in up to four filters using the Astr\textit{O}matic software suite to reduce the pixel to catalog data. The solar system objects within the acquired sources are selected based on a set of criteria depending on their number of observation, motion, and size. The Virtual Observatory SkyBoT tool is used to identify known objects.
}{We observed 20\,221 SSO candidates, with an estimated false-positive content of less than 0.05\,\%. Of these SSO candidates, 53.4\,\% are identified by SkyBoT. KiDS can detect previously unknown SSOs because of its depth and coverage at high ecliptic latitude, including parts of the Southern Hemisphere. Thus we expect the large fraction of the 46.6\,\% of unidentified objects to be truly new SSOs.
}{Our method is applicable to a variety of dithered surveys such as DES, LSST, and Euclid. It offers a quick and easy-to-implement search for solar system objects. SkyBoT can then be used to estimate the completeness of the recovered sample.}%

\maketitle
\section{Introduction}
\begin{figure*}[t]
    \centering
    \includegraphics[width=0.95\textwidth]{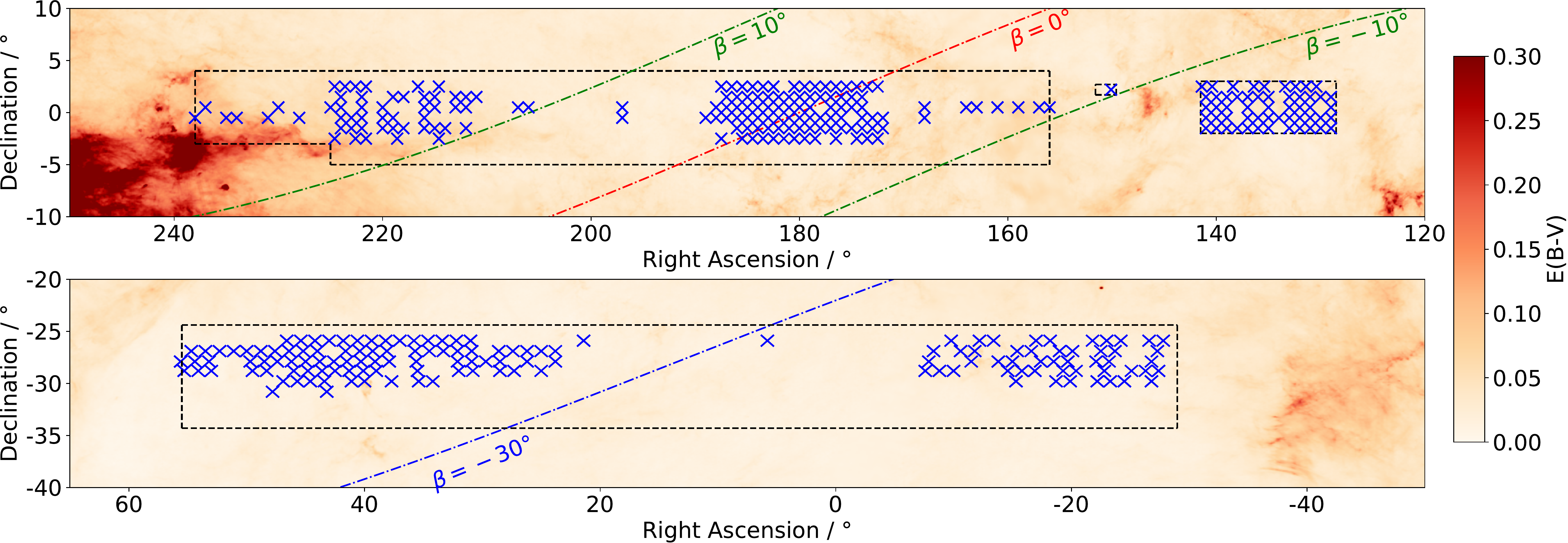}
    \caption{\inl{KiDS field lay-out:} The black, dashed lines mark the area of KiDS-north (top) and KiDS-south (bottom). Each blue cross represents one square-degree field that was searched for solar system objects, in up to four different bands. The ecliptic is shown in the top figure as red, dashed line. The lines of $\pm~10^{\circ}$ ecliptic latitude are shown as green, dashed lines. 66\% of all known solar system objects are within that range of ecliptic latitude. In the bottom figure, the blue, dashed line shows the line of ${-30}^{\circ}$ ecliptic latitude. The background represents the reddening $E(B-V)$ as given by \citet{1998ApJ...500..525S}. After \citet{2017MNRAS.465.1454H}.
}
\label{kids_layout}
\end{figure*}

The study of solar system objects (SSOs), especially the minor bodies, is key in understanding how planetary systems form and evolve. Current populations of comets, asteroids, and trans-neptunian objects are the results of their primordial accretion in the disk and subsequent dynamical evolution, including events such as planetary migrations. As such, the study of the orbits and compositions of minor bodies can provide strong constraints on the formation and the evolution of planetary system \citep{2005Natur.435..459T, 2013-Icarus-226-DeMeo, 2014-Nature-505-DeMeo,AsteroidsIV}. Additionally, it can provide insight into their impact on planetary life, in particular for Earth. In recent years, there has been an increased use of wide-field imaging surveys to discover and characterize serendipitously-observed SSOs \citep[e.g.,][]{2015Icar..261...34V,2016-AA-591-Popescu,2016-Icarus-68-Carry}.\\
The Kilo-Degree Survey (KiDS\footnote{\url{http://kids.strw.leidenuniv.nl}}) is an optical imaging survey of 1500\,${\rm deg}^2$ in an equatorial and a southern patch of the sky, see Fig. \ref{kids_layout}. The areas are imaged in the $u$, $g$, $r$, and $i$ filters using OmegaCAM on the VLT Survey Telescope (VST) at the Paranal Observatory, operated by the European Southern Observatory (ESO) in Chile. The survey is an ESO Public Survey and is described in detail in \citet{2015A&A...582A..62D}.

The KiDS survey products, consisting of calibrated images and catalogs, are released by ESO\footnote{\url{http://eso.org/sci/observing/phase3/data\_releases.html}}. These products, plus all intermediate products, such as calibrated single exposures, and additional survey products, are available via the Astronomical Wide-field Imaging System for Europe \citep[Astro-WISE\footnote{\url{http://www.astro-wise.org}},][]{astro-wise}.

The primary survey design driver was the study of Dark Matter and Dark Energy through weak gravitational lensing \citep{2015MNRAS.454.3500K}. The KiDS survey also offers an opportunity to search for and study SSOs.
The exquisite image quality delivered by OmegaCAM and the VST combined with an observing strategy involving consecutive sets of relatively deep exposures, a large sky coverage, and faint limiting magnitudes make KiDS a good survey to detect and discover SSOs.

KiDS is primarily an extragalactic survey but happens to include about 300\,${\rm deg}^2$ near the ecliptic plane. The remaining 1200\,${\rm deg}^2$ thus offer a chance to characterize the populations of SSOs highly inclined with respect to the ecliptic plane, which are important witnesses of a disturbed dynamical past in the solar system \citep{2017AJ....153..236P}. Beyond the detection and discovery described in the present article, KiDS will allow statistically significant studies of the fundamental properties of these SSOs, in particular of their distributions of velocities, and their inclination with respect to the ecliptic. Lastly, the SSOs can form a contamination for other science cases using the KiDS data \citep[e.g.,][]{2017MNRAS.465.1454H}.

The present article is a pilot study illustrating the strengths of the method we developed to detect SSOs and filter out contaminants.
The method is applied to 65\,\% of the KiDS fields available in the third data release (KiDS DR3) to investigate its suitability for recovering and detecting SSOs. The remaining fields were not searched due to the time constraints of this project and technical difficulties outside the hands of the authors. The subset of KiDS fields we have searched includes 346\,${\rm deg}^2$ of the sky, with 206 located in the KiDS-North field and 140 located in the KiDS-South field, as shown in Fig. \ref{kids_layout}. Of these 346\,${\rm deg}^2$, 18 are intersected by the ecliptic.

We have used the SExtractor\footnote{\url{http://www.astromatic.net/software/sextractor}} and SCAMP\footnote{\url{http://www.astromatic.net/software/scamp}} software packages \citep{sextractor, scamp} to recover the positions, morphologies, and proper motions of sources in the survey images, which are then passed through a cascade of filters based on a priori knowledge of the properties of SSOs.
The method can be applied to any survey employing taking successive, dithered exposures with a CCD array, either in the optical regime (e.g. OmegaCAM at VST, HyperSuprimeCam at Subaru, DECam at Cerro Blanco, VIS on ESA's Euclid mission) or in the near-infrared (e.g. UKIDSS, ESO VISTA, Euclid NISP).

\section{Methods}
\subsection{SSO candidates detection\label{sec:ssocandidates}}

\begin{figure}[t]
    \centering
    \includegraphics[width=0.255\textwidth]{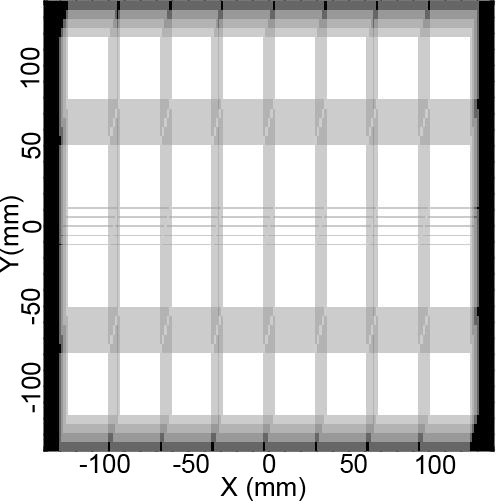}
    \caption{\inl{Dithering pattern:} The OmegaCAM CCD mosaic covers a 1\,${\rm deg}^2$ FoV with gaps between the CCDs. To bridge these gaps, each KiDS field is imaged with 4 (u) or 5 (g, r, i) dithered exposures with a constant diagonal offset in the X and Y directions by 25\,\arcsec~and 85\,\arcsec~respectively. From \url{http://kids.strw.leidenuniv.nl/}.}
    \label{dithering}
\end{figure}

\begin{figure*}[t]
    \centering
    \includegraphics[width=1\textwidth]{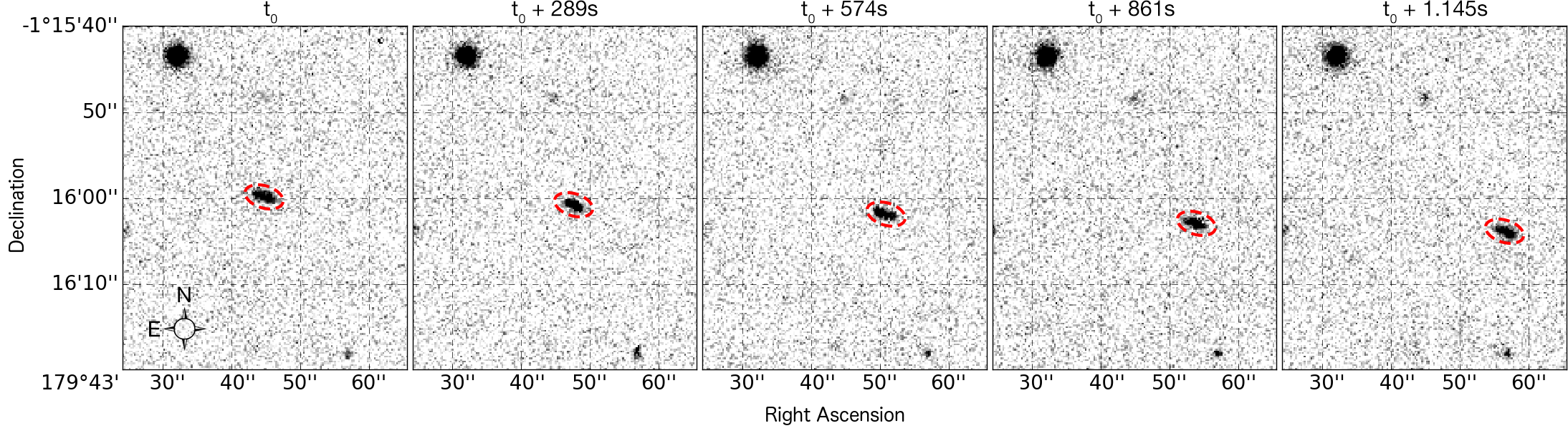}
    \caption{\inl{Example SSO candidate:} An asteroid detected in five different images in the Kilo-Degree Survey. The epoch increases from left to right. The asteroid is indicated by the red ellipse. The total time between the first and the last exposure is 1154\,s. The asteroid has a proper motion of (32.8\,$\pm$\,0.2)\,\arcsec/h. North is up and east is left.}
\label{panorama}
\end{figure*}

We chose to identify asteroids in the KiDS images using the method described in \citet{2013A&A...554A.101B}.
Briefly, the method simply takes advantage of the dithering strategy implemented by KiDS (see Fig.~\ref{dithering}) to search for fast moving objects in the sequence. While stars and galaxies do not move on a timescale of a few minutes typical of dithering sequences, many SSOs display a motion large enough (from a few to many arcseconds per hour) to make their identification possible in proper motion space, as illustrated in Fig.~\ref{panorama}. In each individual image of a sequence, all the sources are detected and their accurate centroid position measured using SExtractor \citep{sextractor}.


The KiDS data products that were used are photometrically and astrometrically calibrated frames (RegriddedScienceFrame in the Astro-WISE data model) along with their respective weight maps. The weight maps produced by the pipeline in Astro-WISE were used to properly take into account the individual pixel statistical properties. The SExtractor configuration and parameter files are provided in Appendix \ref{app:scamp} for the sake of reproducibility. The most important SExtractor settings include:%

\begin{itemize}
\item the detection threshold \verb|DETECT_THRESH|, set to 1.5 standard deviations of the local background in the filtered image.
\item the minimum contrast for deblending \verb|DEBLEND_MINCONT|, set relatively high to 0.05. SSOs  often move so fast that they appear as trails rather than point sources. A low value leads SExtractor to interpret SSO trails as multiple blended objects which then fools the proper motion calculation. On the other hand, a high value that is too high would lead SExtractor to merge neighboring sources into a single detection, and in particular merge SSOs with eventual nearby stars or galaxies. The value was chosen after an extensive heuristic analysis to find the best compromise between the two effects described here.
\end{itemize}

The catalogs are then precisely registered by comparing them to the 2MASS near-infrared catalog \citep{2006AJ....131.1163S} using SCAMP \citep{scamp}, the sources cross-identified within a radius (\verb|CROSSID_RADIUS|) of 10\,\arcsec~and their motion over the duration of the sequence is computed. This cross-match radius was chosen to encompass SSOs with proper motions up to 200~\arcsec/h, therefore including the vast majority of SSOs known to date: this high value corresponds to near-Earth asteroids, while the bulk of known SSOs are in the asteroid main belt, and display motion of typically 20--40\,\arcsec/h. The proper motion of SSOs in the KiDS images is computed by SCAMP, which performs a linear fit of the measured source positions over the observation epochs. A more detailed description of this computation can be found in \citet{2013A&A...554A.101B}.

In addition to the standard centroid positional measurements (\verb|ALPHA_J2000|, \verb|DELTA_J2000|, \verb|THETAWIN_IMAGE| and associated uncertainties \verb|ERRAWIN_IMAGE, ERRBWIN_IMAGE, ERRTHETAWIN_IMAGE|), some important morphological parameters were measured, including the ellipticity (\verb|ELLIPTICITY|), elongation (\verb|ELONGATION|), and semi-major and semi-minor axis length (\verb|AWIN_IMAGE|, \verb|BWIN_IMAGE|). We shall see that these prove useful to filter out contaminants among fast moving SSO candidates.

\begin{figure}[t]
    \centering
    \includegraphics[width=1\columnwidth]{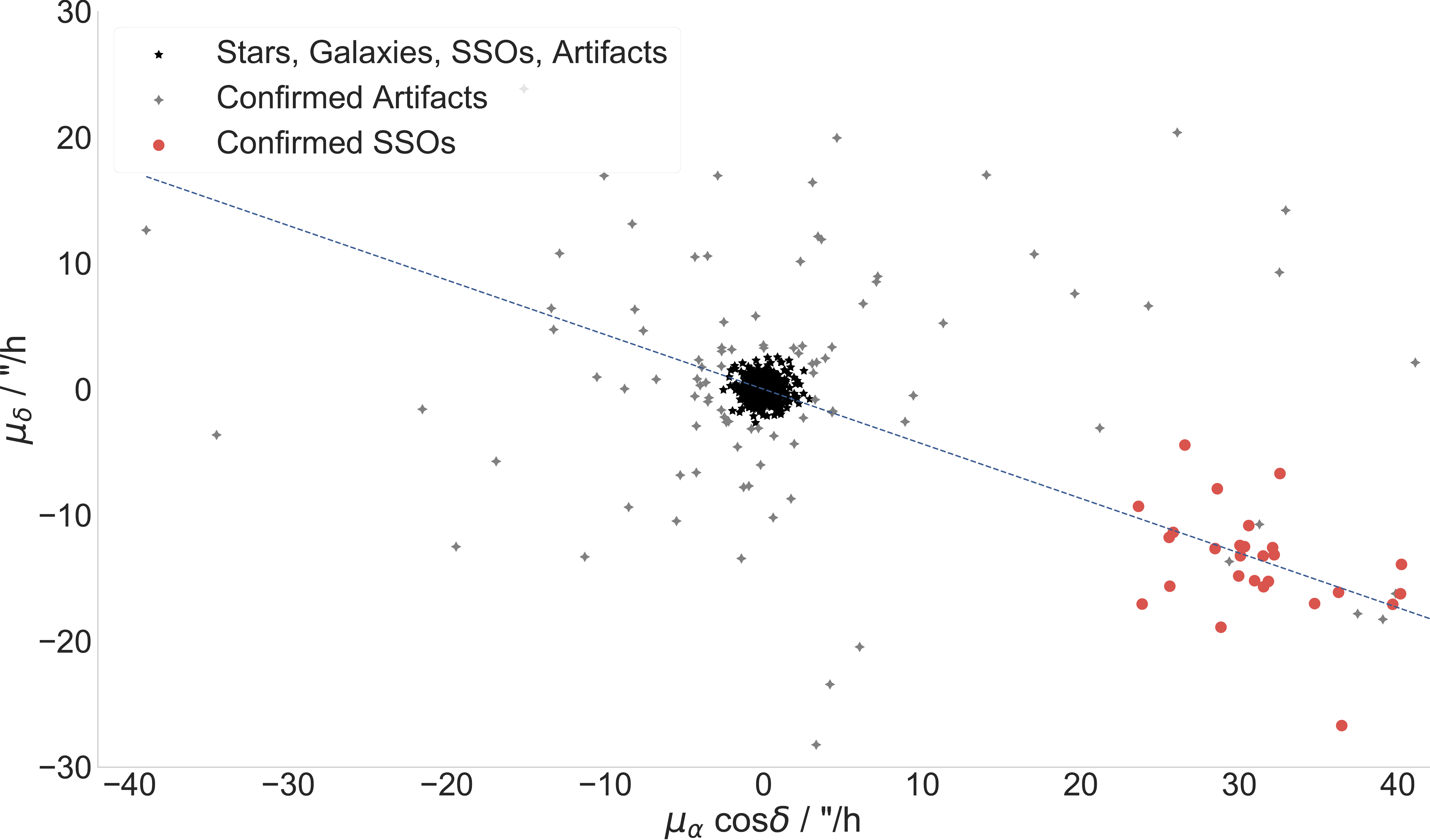}
    \caption{\inl{Proper motion space of SSO candidates:} The black stars represent all the sources below 3\,\arcsec/h proper motion. Objects with larger proper motions were visually inspected and separated into artifacts (gray diamonds) and SSOs (red dots). The direction of the ecliptic is shown by the blue, dashed line. The errors on the proper motion are typically smaller than symbol size.}
\label{pmradec}
\end{figure}

An example of a proper motion space of SSO candidates in a single-band field is shown in Fig. \ref{pmradec}. To show the distribution of SSOs, we visually inspected sources with proper motions larger than 3\,\arcsec/h, a compromise between showing all SSOs and time demand of the visual inspection. We can see a cluster around zero proper motion in right ascension and declination surrounded by a cloud of fast moving sources dispersed in all directions and made up of SSOs and imaging artifacts caused by for example, cosmic rays. In Fig. 4 we indicate objects that were confirmed as SSOs by visual inspection, as well as confirmed artifacts. On the figure we can also see the cluster around zero proper motion which are mostly stars and galaxies, however, some artifacts and slow-moving SSOs (e.g. trans-neptunian objects) fall into the same region. The direction of the ecliptic is indicated on the figure. For clarity, only a sixth of the 17\,719 sources in this single-band field are shown.
\begin{table*}[th]
\centering
\caption{\inl{KiDS survey observing strategy:} Exposure times in seconds, number of total exposures, the mean sequence duration in seconds, the full-width half-maximum of the point spread function in arcseconds, and the limiting magnitude for the different filters. The last two columns give an estimate of the theoretical lower proper motion limit derived by the sequence duration and residual dispersion, and the upper proper motion limit given by the cross-matching radius. After \citet{2015A&A...582A..62D}.}
\label{filter}
    \begin{tabular}{llllllll} \toprule
 Filter & Exposure time& Exposures & Sequence duration& PSF FWHM& Mag limit & Lower PM limit& Upper PM limit\\
 & (s) & & (s) & (\arcsec) & (AB 5$\sigma$ 2") & (\arcsec/h)& (\arcsec/h) \\
 \midrule
  u   & 250   & 4  & 1176 & 1.1 & 23.8-24.4 & 0.05 - 0.08 & 160\\
  g   & 180   & 5  & 1120 & 0.9 & 24.6-25.4 & 0.05 - 0.08 & 200\\
  r   & 360  & 5  & 2020 & 0.8  & 24.6-25.3 & 0.03 - 0.04 & 100\\
  i   & 240  & 5  & 1420 &  1.1 & 22.9 - 24.4 & 0.04 - 0.06 & 167\\ \bottomrule
 \end{tabular}
\end{table*}
\subsection{Detection envelope}
A detailed analysis of the parameter space probed by the present study would require extensive simulations that go well beyond the scope of this article. In this section, we determine the detectability of moving objects in terms of proper motion and luminosity based on the general KiDS survey properties described in \citet{2015A&A...582A..62D} and summarized in Table~\ref{filter}.
The proper motion domain probed by our analysis mostly depends on the precision of the centroid position measurement and dithering strategy of the survey. Given the excellent optical quality of the VST and seeing requirements of the survey (better than 1.1~\arcsec, see Table~\ref{filter}), the precision of the centroid measurement for high signal-to-noise ratio (photon-noise limited) detections over such short exposure times is largely dominated by the error introduced by the atmospheric turbulence. For high-signal-to-noise ratio sources we find a typical residual dispersion in our astrometric analysis of only 15--25~mas depending on the seeing and band (being worst in the u band). These numbers go up to 20--30~mas for low signal-to-noise ratio sources ($\ge$5$\sigma$).

The individual exposure times and corresponding 5-$\sigma$ limits of sensitivity are given in Table~\ref{filter}.
The observing strategy includes the acquisition of five consecutive dithered images in the $g$, $r$, and $i$ filters and four in the $u$ filter. The dither pattern is a constant diagonal offset in the X and Y directions by 25\,\arcsec~and 85\,\arcsec~respectively (see Fig.~\ref{dithering}). The total dither sequence duration typically lasts between 1120--2020~s depending on the filter and is made up mostly of the science exposure, read-out, telescope offset and, in some cases, filter change.
The same field is not necessarily observed in the four filters consecutively, and several days, weeks or even months can pass until a given field is observed in the next filter. As a consequence, we chose to analyze the four filters completely independently. Some asteroids might therefore appear several times in our catalog without being cross-identified.

An estimate of the upper and lower limit of the proper motion domain probed by our analysis is given in Table\,\ref{filter}. The upper limit is given by the cross-matching radius set in the SCAMP configuration, while the theoretical lower limit is derived by dividing the residual dispersion in the astrometric analysis by the mean sequence duration of each filter. Also given in Table\,\ref{filter} is the limiting magnitude for individual exposures in each filter, which inherently limits the magnitude of the SSOs that we can recover. The magnitude limits vary from exposure to exposure due to differences in seeing conditions and sky brightness.
\subsection{SSO candidates selection}\label{contaminants}
\begin{figure}[t]
    \centering
    \includegraphics[width=\columnwidth]{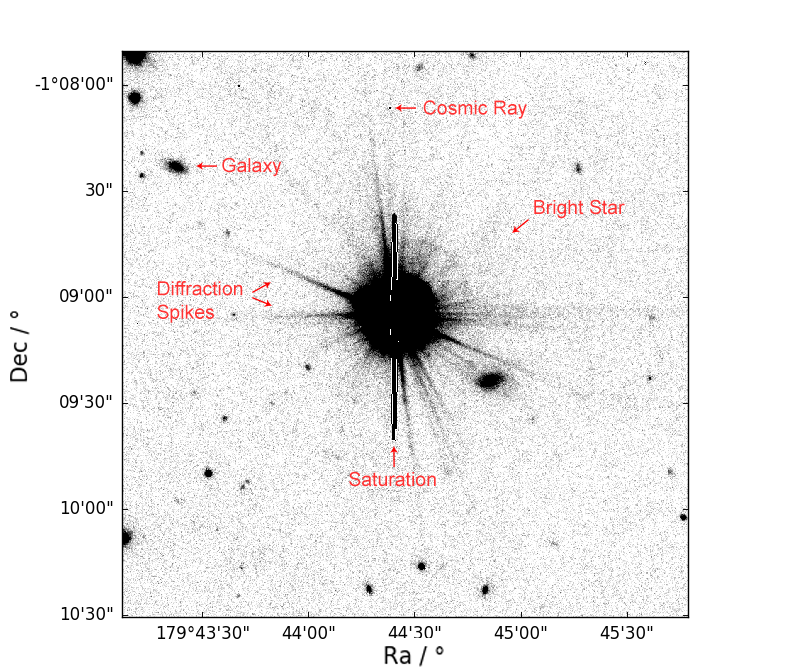}
    \caption{\inl{Collection of artifacts around bright star:} Besides cosmic rays, bright stars are the main source of sample contaminants, as they introduce saturation effects, diffraction spikes, and halos in the images. Also visible are less bright stars and galaxies.}
    \label{artifact}
\end{figure}

\begin{figure}[t]%
\centering
\subfloat{\label{wls_ra}\includegraphics[width=0.98\columnwidth]{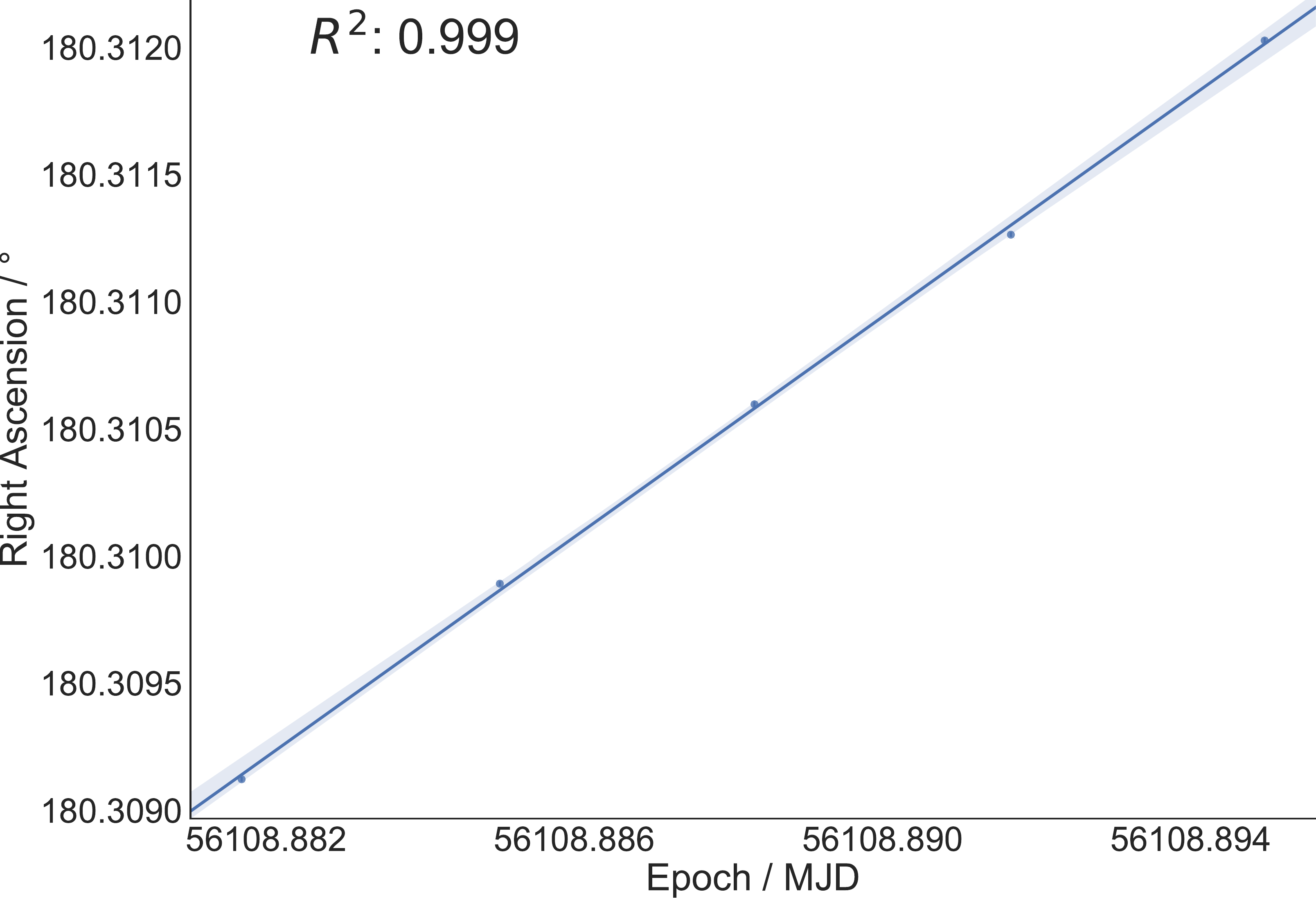}}\\
\subfloat{\label{wls_dec}\includegraphics[width=0.98\columnwidth]{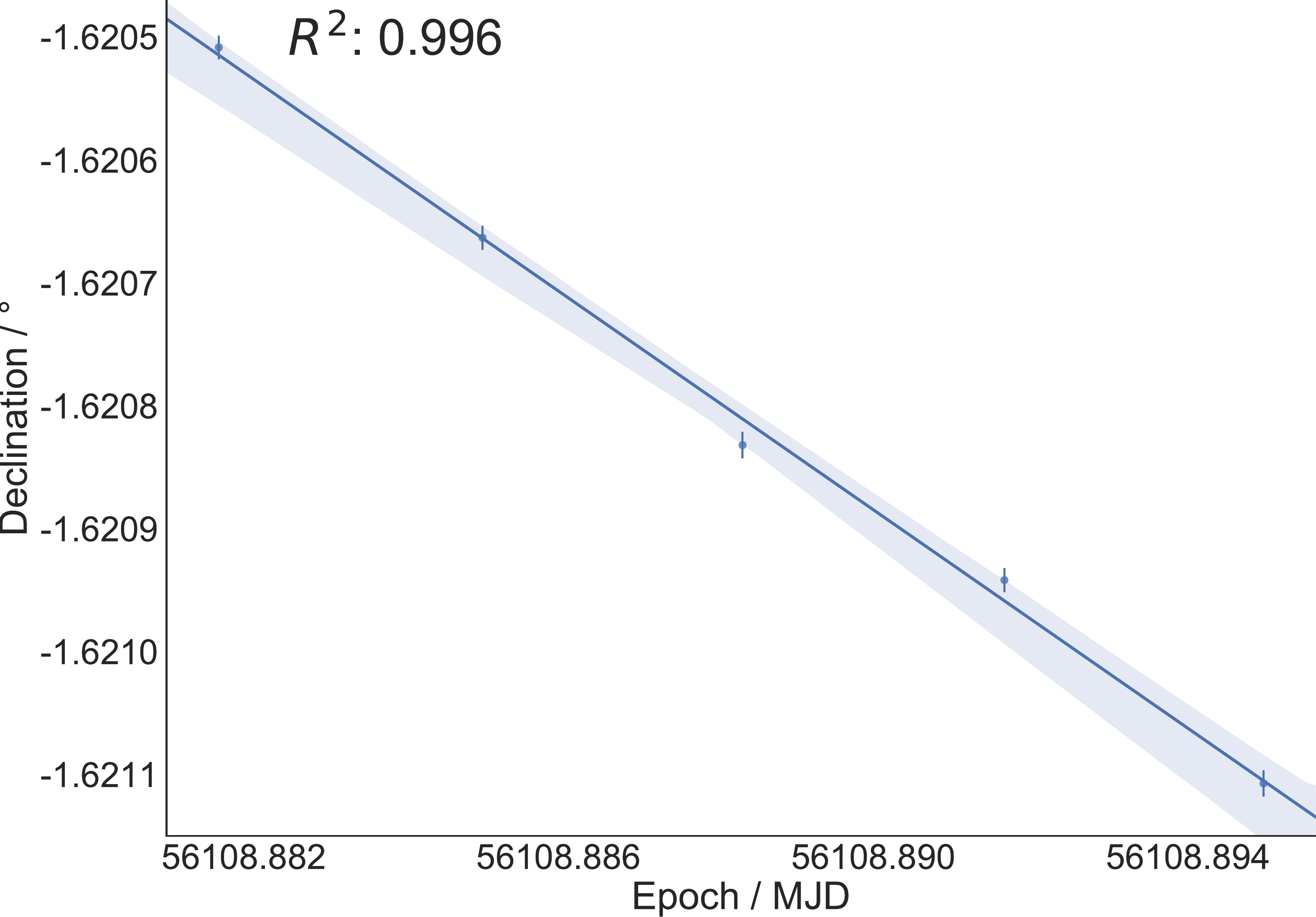}}%
\caption{Motion of an SSO: A linear weighted-least-squares linear fit of the motion of an SSO in right ascension (top) and declination (bottom). The blue dots represent the positions and the solid lines the linear fits. The shaded blue areas mark the 95\,\% confidence interval. The positional errors are the errors on the centroid computation performed by SExtractor. In the upper plot, the errors on the right ascension are smaller than the symbol sizes.}\label{wls}
\end{figure}

Once the proper motion of all the sources had been measured in a dithered sequence, SSO candidates were selected following a procedure that tries to find the best compromise between the contamination and completeness of the sample. Contamination by various kinds of artifacts is inevitable in such a large dataset. An extensive visual inspection of a sample dataset covering 3~deg$^2$ in the $i$ band near the ecliptic plane allowed us to define a "bestiary" of the most common contaminants in the sample before applying the filtering cascade:
\begin{itemize}
\item cosmic rays can fall by chance within a cross-match radius in two or more consecutive frames and mimic a moving source;
\item bright star halos and diffraction patterns can be extracted as sources by SExtractor. As the corresponding peaks often move from frame to frame, they can also mimic a fast moving source;
\item stars and galaxies contaminate the sample. Because outer solar system objects, such as Kuiper-belt objects, have very low proper motions (below 1--2\arcsec/h), filtering with a hard cut on the smallest proper motions is undesirable here. We instead discriminated between fixed-coordinates objects (stars, galaxies) and SSOs by checking the linearity of their motion across different frames (see Fig. \ref{panorama})

\end{itemize}
A typical collection of artifacts is shown in Fig. \ref{artifact}. Visible are galaxies, stars, a saturation trail, and diffraction spikes around the star in the center. While many imaging artifacts like dead pixels are already masked by the weight images provided by Astro-WISE and other contaminants like stars and galaxies can be filtered efficiently with the cascade of filters described below, the diffraction spikes will prove to be the most difficult artifacts to filter as they mimic the trails of SSOs in motion and size. SExtractor detects different parts of a diffraction spike as objects, which are then later cross-identified by SCAMP. The uniform size and the apparent motion introduced by the different orientation of the diffraction spikes in each exposure make these objects appear like SSOs.
Using this library of contaminants, we then performed an extensive heuristic analysis to iteratively define the following cascade of filters:

\begin{enumerate}
\item Number of detections: Candidates are required to have been detected at least in four different individual images. This is particularly efficient to reject cosmic rays. As a consequence, SSOs detected in only two or three images are missed by our analysis. Later in this paper we investigate how including objects with three detections changes the sample size and purity.
\item Linear motion: Over the maximum duration of the dither sequence, the motion of SSOs can be considered linear. We found that many artifacts made of randomly coincident cosmic rays, hot pixels, halos, diffraction spikes, stars, and galaxies displayed non-linear motions. A goodness-of-linear-fit analysis is therefore performed using the $R^2$ parameter of the weighted-least-squares linear fit in both right ascension and declination over time as diagnostic. A moving source is considered a good candidate if both $R^2$ parameters are larger than 0.95, and rejected otherwise. Figure~\ref{wls} illustrates the procedure.
\item Proper motion: The upper and lower proper motion limits for each filter as given in Table \ref{filter} are applied to the remaining candidates. Additionally, objects with a relative error on the proper motion larger than 5\% are filtered out. This relative error effectively increases the lower limit on the proper motion of the SSO. Depending on the dispersion in the image, the minimum error on the proper motions are between 0.03\,\arcsec/h and 0.08\,\arcsec/h. Therefore, depending on the filter the objects were observed in, the filter on the relative error increases the lower proper motion limits to 0.6\,\arcsec/h to 1.0\,\arcsec/h.
\item Trail size: Because the individual exposure times in a dither sequence are equal, the size of the SSO trails is expected to be constant within the error. We therefore check that the semi-major and semi-minor axis of the SSOs trails measured by SExtractor are constant in the images. We do that simply by checking the $R^2$ parameter of the weighted-least-squares constant fits on the semi-major and -minor axis over time. A moving source is considered a good candidate if both $R^2$ parameters are larger than 0.95, and rejected otherwise. This filter is highly efficient at rejecting the remaining cosmic rays and to a lower extent bright star halos.
\item Trail size distribution: This step consists in filtering out objects with sizes significantly larger than 95\% of the remaining candidate population in the current single-band field. It was motivated by the fact that most remaining contaminants at that stage were part of ghosts and halos of bright stars which typically are highly extended shapes.
\item Proximity to bright stars: This filter was added retroactively after applying the pipeline on the whole set of available KiDS images. We find that more than 75\,\% of the artifacts in the whole sample of SSO candidates were introduced by bright stars. The sample is therefore cross-matched with the HYG database\footnote{\url{http://www.astronexus.com/hyg}}, which contains all stars in the Hipparcos, Yale Bright Star, and Gliese catalogs. The cross-match radius of 300\,\arcsec is chosen by looking at typical sizes of clusters of candidates around stars. Candidates within this radius are rejected.
\end{enumerate}
The parameters of the first five filter were found iteratively by running the pipeline on the 3~deg$^2$ test fields and evaluating the purity and the completeness of the output sample. We set the parameters more toward a pure rather than a complete sample. After visual inspection in the test-bench field, we find that the filter cascade eliminates of 99.957\,\% of contaminants (four out of 92555 contaminants left) while recovering 84.6\,\% of the SSO present in the images (66 out of 76). Applying the final filter based on the proximity to bright stars removes the remaining four contaminants.

The ten missing SSOs were not recovered for one of three reasons:
\begin{itemize}
    \item They passed close to a bright object and were associated by SCAMP with this object. Therefore, they were rejected by either the linear motion filter or the trail size filter as neither size nor apparent motion fit the expected behavior.
    \item Due to the dithering strategy of KiDS, SSOs on the edge of the imaged area are sometimes out of frame for one exposure. These objects are rejected by the number of detections filter, as SCAMP does not link all four detections, but rather as two sets of two.
    \item Significant magnitude variations within the SSO trails during the exposure. SExtractor will then interpret the trail as two different objects.

\end{itemize}
As mentioned in Section~\ref{sec:ssocandidates}, we reduced the number of these cases by adjusting SCAMP and SExtractor parameters, however, not all SSOs could be recovered.

\begin{table}[t]
\centering
\caption{Results: For each filter, the number of fields searched, and the number of SSO candidates in all fields.}
\label{general_results}
    \begin{tabular}{lll} \toprule
 Filter & Number of fields & Number of SSOs\\ \midrule %
  u   & 267   & 879 \\%
  g   & 297   & 8\,124 \\%
  r   & 308  & 7\,362 \\  %
  i   & 270  & 3\,856   \\ \bottomrule  %
 \end{tabular}
\end{table}
\subsection{Comparison to the list of known SSOs\label{sec:skybot}}
We cross-matched the result of our analysis to the known list of SSOs by means of the Virtual Observatory SkyBoT utility \citep{skybot}. SkyBoT provides a fast and simple cone-search method to list all known SSOs within a given field of view at a given epoch. For that, it weekly precomputes ephemerides of all known SSOs, based on their osculating elements computed at the Lowell observatory \citep[ASTORB database,][]{1993-LPI-Bowell} and IMCCE \citep[cometpro database,][]{1996-IAUS-Rocher}, and stores them in a hierarchical database supported by nodes based on geocentric equatorial coordinates. The accuracy on the positions is thus directly dependent on the accuracy of the input osculating elements, and while better than 1\arcsec for 68\% of SSOs, it can be extremely poor for badly constrained orbits.

For each single-band field, we looked for counterparts to our final sample of SSO candidates in the SkyBoT outputs within a radius of 10\arcsec. This search radius was chosen as a conservative compromise to avoid mis-associations while recovering as many SSOs as possible. When a match was found, the SSO name and predicted magnitude provided by SkyBoT were added to our database.

\section{Results}\label{sec:results}

In the present analysis, 346\,${\rm deg}^2$ of the sky were searched for SSOs in up to four bands. In total, 1\,142 single-band fields were analyzed, corresponding to 65\,\% of KiDS survey Data Release 3, as shown in Fig.~\ref{kids_layout}. We report the finding of 20\,221 SSO candidates, with four to five observations each. These candidates represent 0.06\,\% of all the initial objects detected in the dataset.

Table~\ref{general_results} shows that most candidates were recovered in the $g$ and the $r$ band. This is expected for objects seen in reflected sunlight. Inside a dithering sequence, 52.3\,\% of all SSOs were observed five times and the remaining 47.7\%  four times (we rejected candidates with only two or three detections, see above).

\subsection{Purity}

To estimate the false-positive rate of our method, we performed extensive studies of the sample and different subsamples. A false-positive consists of an object passing the chain of analysis that is not an SSO. We quantify this number by visually confirming subsets of our sample and using the false-positive content as an estimator to infer on the statistics of the whole population.\\
It is more efficient to study the distribution of artifacts on a sample with a larger artifacts content. We therefore start with a larger sample, containing not only the 20\,221 SSO candidates with four or five detections each, but also candidates with three detections which pass the filter cascade. Also included are all objects within the cross-match radius around bright stars.\\

The extended sample contains 28\,290 candidates. We categorized the objects of 103 randomly selected fields of this larger sample into "SSO" and "artifact". After visually confirming the objects in these 103 fields, the false-positive rate of the extended sample is estimated to be 6.8$\,\pm\,$0.5\,\%.\\
The vast majority of the contaminants found this way could be put into two, overlapping groups. Most artifacts were observed three times only, and a large part were reflection ghosts. These artifacts were introduced into the images by bright stars and perfectly mimic SSOs in both appearance and linear motion. \\
Removing all objects which were only observed in three images from the sample decreases the sample size by 25.5\,\%, down to 21\,072 candidates. Looking at 3\,000 randomly selected objects, we find nine artifacts. The false-positive content of this subsample is therefore 0.3$\,\pm\,$0.2\,\%, where the uncertainty is a 1$\sigma$ limit.\\
By studying the spatial distribution of artifacts, we find that they cluster around bright stars with typical cluster radii of the order of 100\arcsec. Cross-matching the sample of all candidates with three to five detections each with the HYG database using a cross-matching radius of 300\arcsec revealed 2404 matches. After inspecting 1000 objects, the false-positive content of this subsample was estimated to be 60.9$\,\pm\,$1.2\,\%. This subsample therefore contains about 76\,\% of all artifacts in the sample of 28\,290 candidates. Subtracting all cross-matched objects within 300\,\arcsec of stars reduces the sample size by 8.5\,\% to 25\,886 while reducing the false-positive content to 1.8$\,\pm\,$0.6\,\%.\\
As the subsample of candidates close to stars also contains the majority of artifacts with four or five observations each, excluding these candidates also decreases the contamination in the sample of objects with four or five observations. The sample size decreases from 21\,072 by 4\,\% down to 20\,221, while the false-positive rate decreases to less than 0.05\,\%, where we used 2$\sigma$-confidence intervals to arrive at the upper limit.\\
Table\,\ref{tab:fpr} sums up the results of this study. Simple cuts decrease the false-positive content by two orders of magnitudes, while excluding 28.9\,\% of the sample. A more sophisticated selection of the exclusion regions around bright stars could improve the method even further. The location and extension of reflection ghosts in the image introduced by the stars can be calculated. Knowing these properties, one could exclude the artifacts while keeping a larger fraction of SSOs. However, this requires extensive calculations outside the scope of this pilot study.
\begin{table}[ht]
\centering
\caption{Sample size and false-positive rate: The false-positive rate (FPR) and the sample size for different samples.}
\label{tab:fpr}
    \begin{tabular}{lll} \toprule
 Sample & Sample size & FPR / $\%$\\ \midrule
  3-5 detections   & 28\,290   & 6.8$\,\pm\,$0.5 \\
  Bright-star regions excl. & 25\,886  & 1.8$\,\pm\,$0.6 \\
    3-detection candidates excl.   & 21\,072  & 0.3$\,\pm\,$0.2 \\  %
  Both excluded   & 20\,221  & $\leq 0.05$   \\ \bottomrule  %
 \end{tabular}
\end{table}
\subsection{Completeness}

As mentioned in Secttion\,\ref{contaminants}, we placed emphasis on a pure sample rather than a complete one. A comprehensive study of the completeness of our sample as done above for the purity by for example injecting fake SSOs into the images is outside the scope of this work. However, the comparison of the sample of SSOs detected in KiDS with the current known population of SSOs (see Section~\ref{sec:skybot}) provides an effective way to assess the completeness and validity of our method.

 Of the 20\,221 SSO candidates detected in this study, 10\,793 (53.4\,\%) have a counterpart predicted by SkyBoT within 10\arcsec and are identified. Conversely, 46.6\,\% of our sample thus are observations of potentially new discoveries of SSOs by KiDS.
The other way around, of the 34\,023 SSOs that SkyBot predicts to be in the searched fields at the respective epochs, 31.7\,\% are found in the KiDS images. We attributed the unmatched 68.3\,\% of SkyBoT SSOs to several factors, including
\begin{itemize}
\item the orbital uncertainty of 21\,\% of the SSOs predicted in the fields of view are larger than 10\,\arcsec, as given by their current ephemeris uncertainty \citep[CEU, see][]{1993-LPI-Bowell};
\item the SSOs are too faint to appear in the individual KiDS images. Visual inspection of the predicted positions of some unmatched SSOs in the KiDS images confirms this assumption;
\item incompleteness of our sample. SSOs which were only detected up to three times in the images were discarded. Bright stars and other artifacts can mask SSOs and prevent the recovery;
\item fainter SSOs usually have larger uncertainties on their orbits.
\end{itemize}

\begin{figure}[t]
    \centering
    \includegraphics[width=\columnwidth]{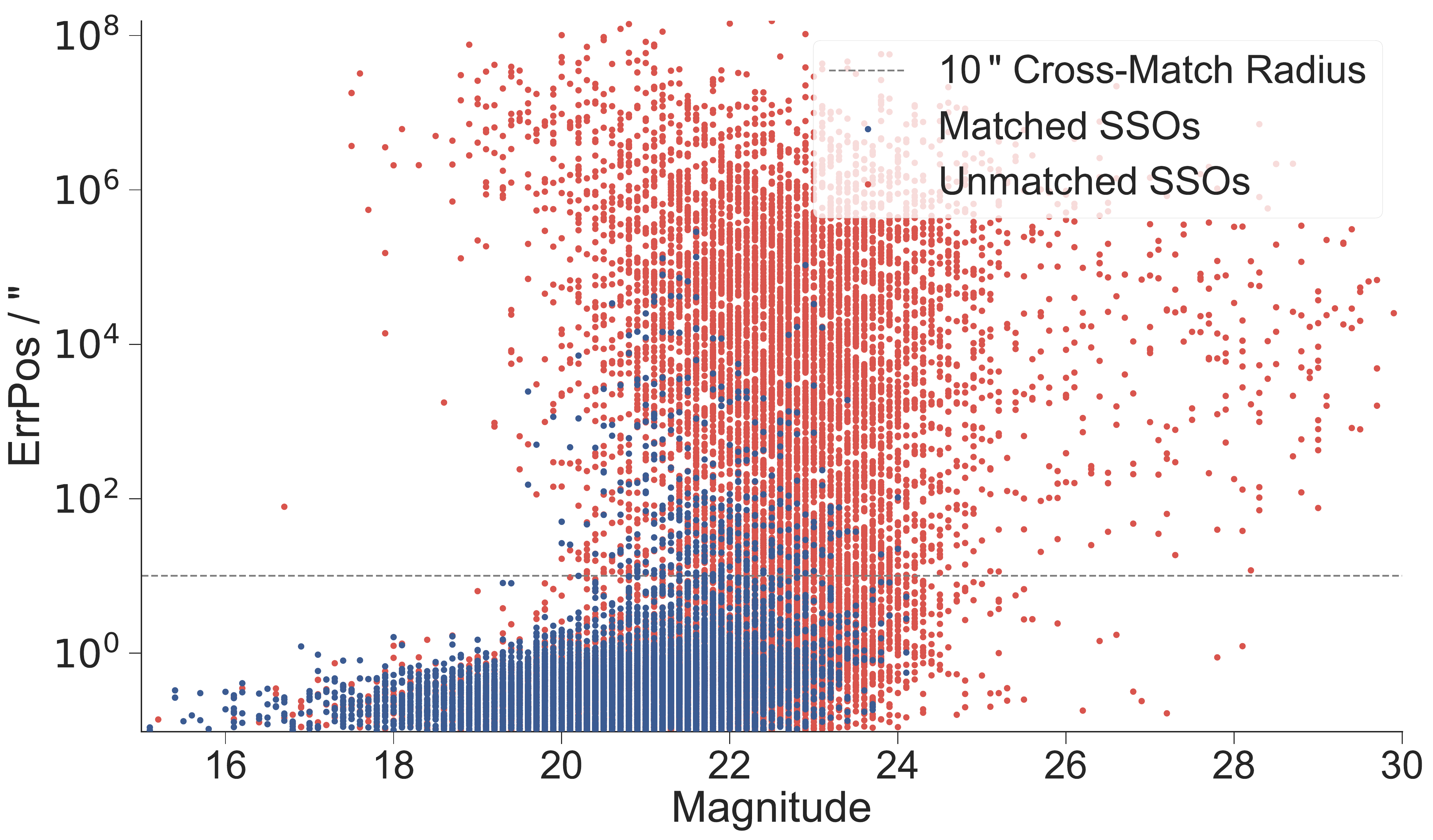}
    \caption{\inl{Predicted magnitudes and positional errors of matched and unmatched SkyBoT SSOs}: The blue dots show the magnitude over the positional uncertainty as given by SkyBoT for SSOs that were cross-matched with a KiDS SSO. The red dots show SkyBoT SSOs that were not found. The gray, dashed line indicates the 10\,\arcsec cross-matching radius. Not shown are objects with positional errors smaller than 0.1\,\arcsec. The column pattern in the data is caused by the limited resolution of the SkyBoT magnitude prediction.}
    \label{mag_skybot}
\end{figure}

The last point is shown in Fig.\,\ref{mag_skybot}, where the error on the predicted position of an SSO is plotted against the predicted magnitude. The matched and unmatched SkyBoT SSOs are shown as blue and red points respectively. The trend for a large positional uncertainty toward fainter magnitudes is clearly visible. The gray, dashed line indicates the 10\,\arcsec cross-match radius we used. The SkyBoT SSOs that were found in KiDS are mostly below this line, while a small fraction extends to up to tens of degrees in positional error. The ten matched SSOs with the largest orbit uncertainties were visually confirmed to be SSOs.

The excluded 8069 SSO candidates with only three observations or within 300\arcsec of stars have a SkyBoT match ratio of 32.0\%, the extended sample of 28\,290 SSO candidates with three to five observations each has a match ratio of 47.1\%.

\begin{figure}[t]
    \centering
    \includegraphics[width=\columnwidth]{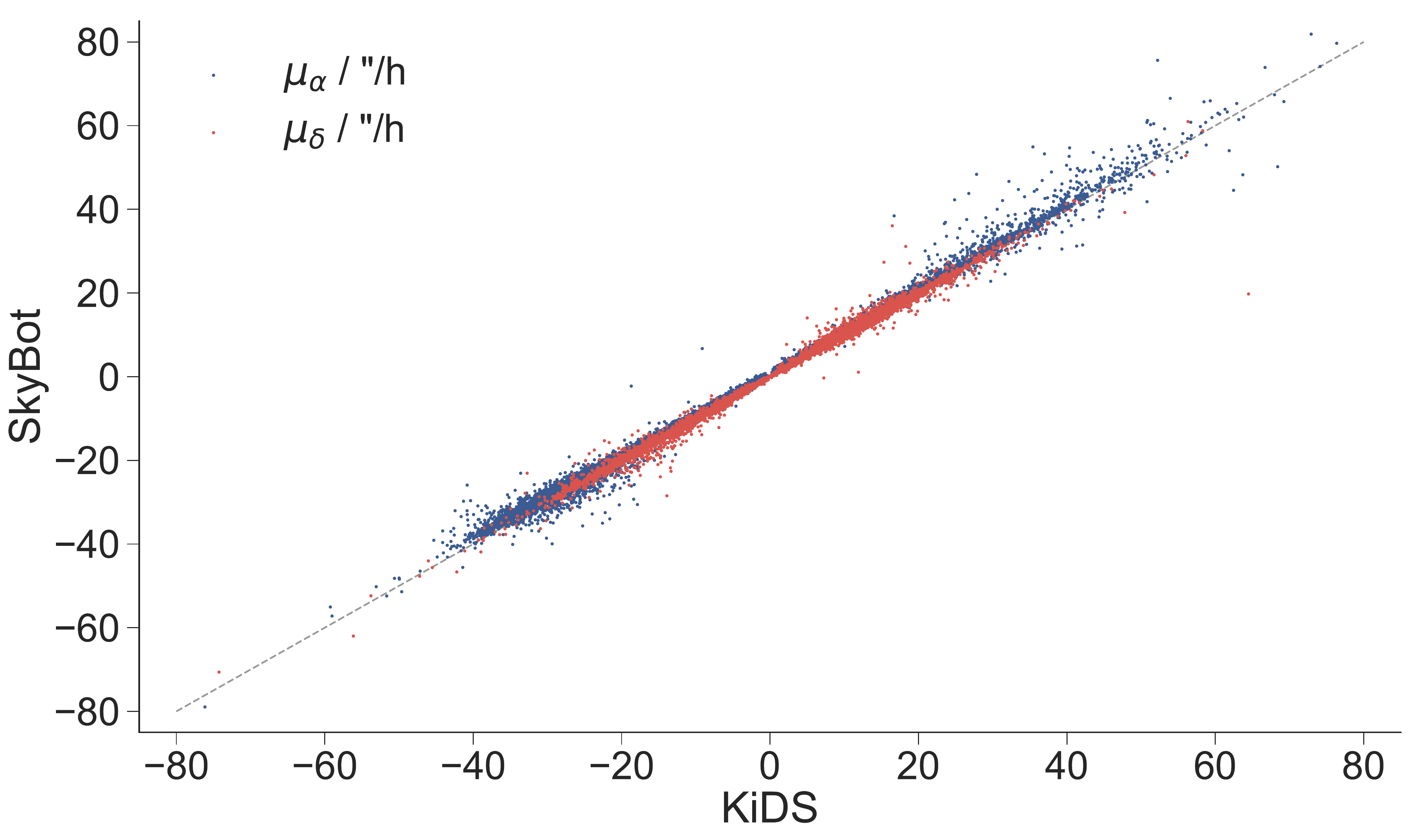}
    \caption{\inl{Proper motion space of SSOs in KiDS and their matches in SkyBot:} The predicted and the observed proper motion values of all cross-matched SSOs in right ascension (blue) and declination (red). The dashed, gray line marks where the predicted and observed values are equal.}
    \label{pm_kids_skybot}
\end{figure}

In Fig.\,\ref{pm_kids_skybot} we compare the proper motion values in right ascension and declination of cross-matched SSOs as given in SkyBoT and as derived from the KiDS images. Points close to the bisecting line through the origin shown in gray show a good agreement between the predicted values by SkyBoT and the observed values in KiDS. The number of outliers increases to higher absolute values, especially for the proper motion in right ascension. Overall, there is a good agreement between the predicted and the observed proper motion values of KiDS and SkyBoT.
\\

All observations of candidates with three to five detections each have been reported to the Minor Planet Center\footnote{\url{http://minorplanetcenter.net}} (MPC). The observations will be used to refine the orbit calculations of known SSOs, which will be especially useful for the SSOs with large positional uncertainties, shown in Fig.\,\ref{mag_skybot}. New designations of SSOs are not expected, as the KiDs observation strategy did not allow for observation of objects over consecutive nights. While we likely did observe SSOs several times, connecting the observations of SSOs is not possible due to the small arcs observed. Extrapolating the positions over the typical time-span between the observations of neighboring fields results in too large errors. Other surveys may prove to be more suited for this task.

The MPC confirmed our observation of 12\,170 SSOs\footnote{Refer to MPC 105287-105576}. This number is smaller than the 13\,335 SkyBoT matches that we get for the sample of 28\,290 candidates as the MPC uses a smaller cross-match radius of 2\,\arcsec. Of 29 cross-matched SSOs with positional uncertainties larger than 10$^4$\arcsec, the MPC accepted 17. Looking at the ratios of the proper motions predicted by SkyBoT and the proper motions measured in KiDS, 25 objects with large positional uncertainties display ratios close to one (refer to Fig.\,\ref{pm_kids_skybot}), while four show large deviations between predicted and measured velocity. This indicates that these four objects were by chance positionally close to where another objects was expected. One coincidental match is accepted by the MPC, three are rejected. Of the 25 remaining objects, nine are rejected even though the proper motions match well. For all 25 objects, one can therefore say that the ephemeris uncertainties are largely overestimated.

Two tables containing the raw data like right ascension, declination, epoch of observation, and derived values such as the proper motion value and SkyBoT association are made available by the Centre de Données astronomiques de Strasbourg\footnote{The data is only available in electronic form at the CDS via anonymous ftp to \mbox{cdsarc.u-strasbg.fr} (130.79.128.5) or via \url{http://cdsweb.u-strasbg.fr/cgi-bin/qcat?J/A+A/}.} (CDS). One table consists of the sample of 20\,221 SSO candidates with an estimated contamination of below 0.05\,\%, and the other contains the sample of the 8\,069 SSO candidates with three detections or within 300\,\arcsec of stars, with a contamination of approximately 24\,\%.

\section{Discussion}\label{sec:discussion}
With the limitations on the completeness and contamination of our sample in mind, we tentatively interpret a number of statistical properties of the sample of SSO found in our study. The results presented in this section are based on the sample with the lowest false-positive content, 20\,221 SSO candidates with four or five observations each and not within 300\,\arcsec of stars.

\subsection{Properties of detected SSOs}

\begin{figure}[t]
    \centering
    \includegraphics[width=1\columnwidth]{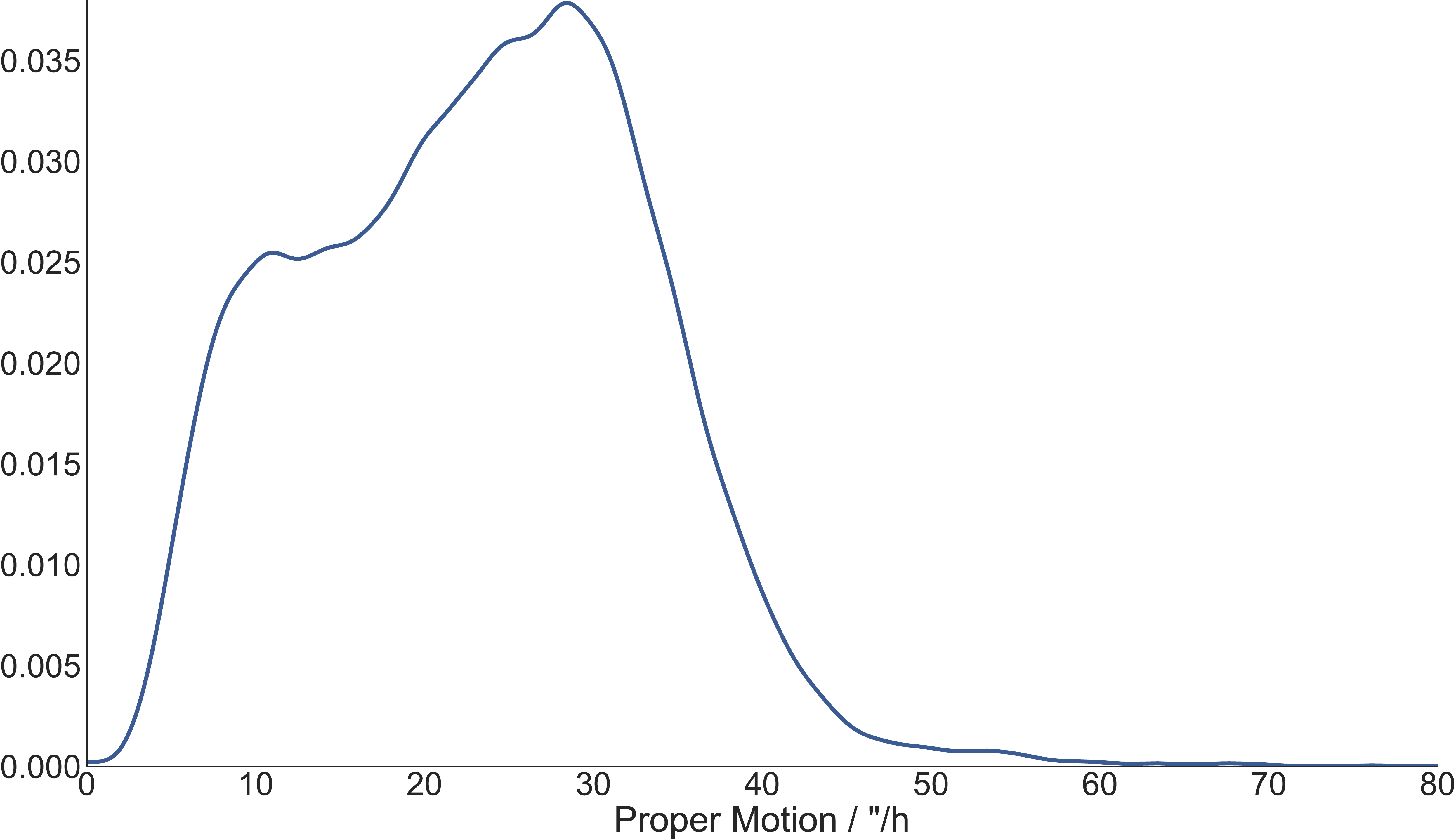}
    \caption{\inl{KDE of proper motion}: The normalized distribution of the proper motions of all recovered SSO candidates in the KiDS images. Not shown is the tail of objects faster than 80\arcsec/h, with the highest proper motion measured at 97\arcsec/h. Candidates with proper motions larger than 80\arcsec/h make up less than 0.5\,\% of the sample.}

    \label{pm_kde}
\end{figure}

\begin{figure}
    \centering
    \includegraphics[width=\columnwidth]{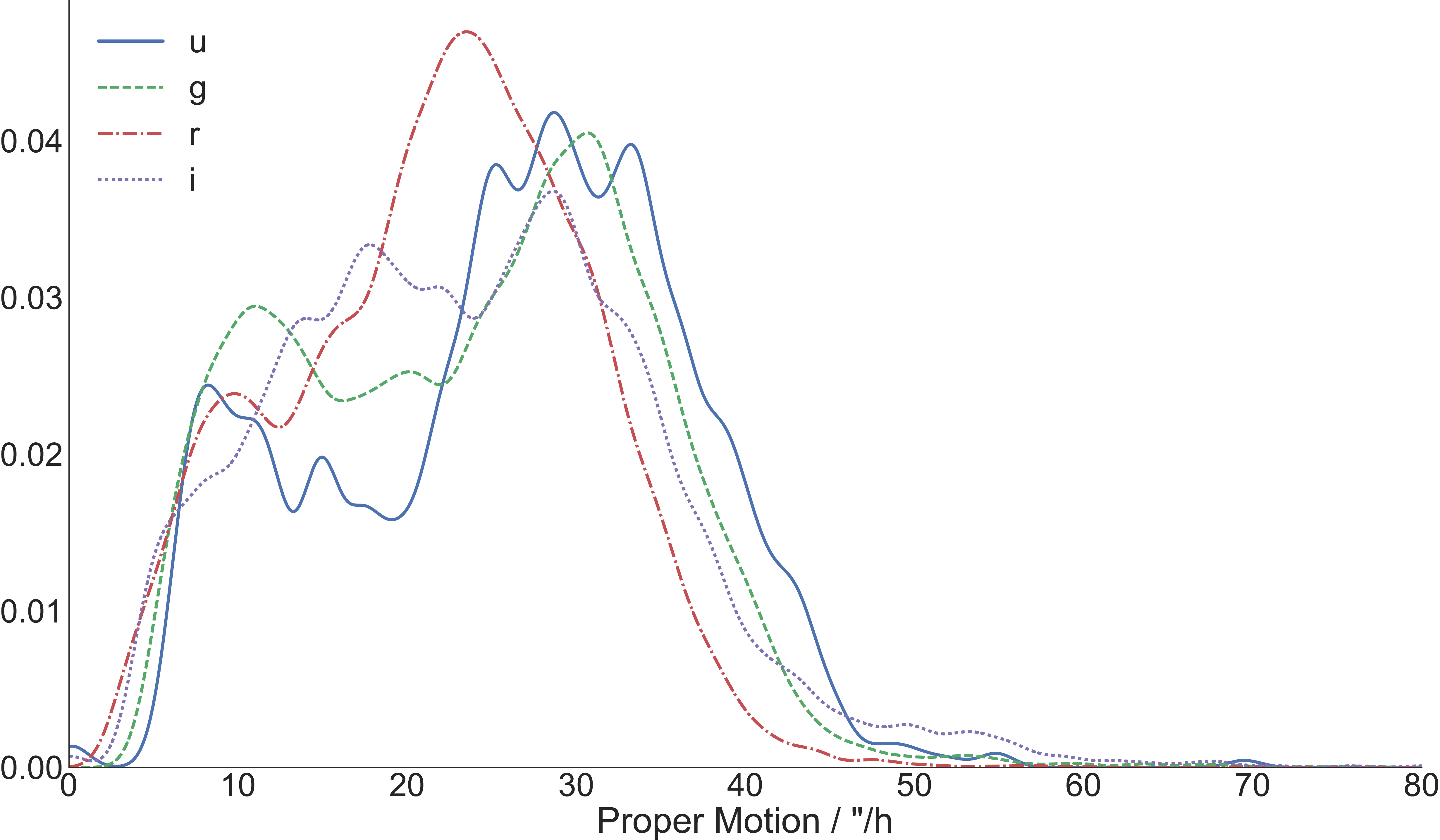}
    \caption{\inl{KDE of proper motion in different bands}: Same as Fig.\,\ref{pm_kde}, except that the SSOs have been split up into sub-samples for the bands they were observed in.}
        \label{pm_kde_ugri}
\end{figure}

We present in Figs.~\ref{pm_kde}, \ref{pm_kde_ugri}, \ref{mag_kde_ugri}, and \ref{ecliptic_latitude} the kernel density estimation (KDE) of the SSOs' proper motion (all together and split by filters), apparent magnitude, and ecliptic latitude. The latter is compared to the KDE of the ecliptic latitude of known SSOs.

The proper motion distribution in Fig.\,\ref{pm_kde} shows that the majority of the recovered SSOs can be identified as main asteroid belt objects, which are characterized by proper motions between 20\,\arcsec/\,h and 40\,\arcsec/\,h. The proper motion space spans from the lower limit of 0.6\,\arcsec/\,h up to 97\,\arcsec/\,h, showing that our method is capable of detecting objects from fast-moving near-Earth to slower trans-neptunian objects in the outer solar system.

\begin{figure}[t]
    \centering
    \includegraphics[width=\columnwidth]{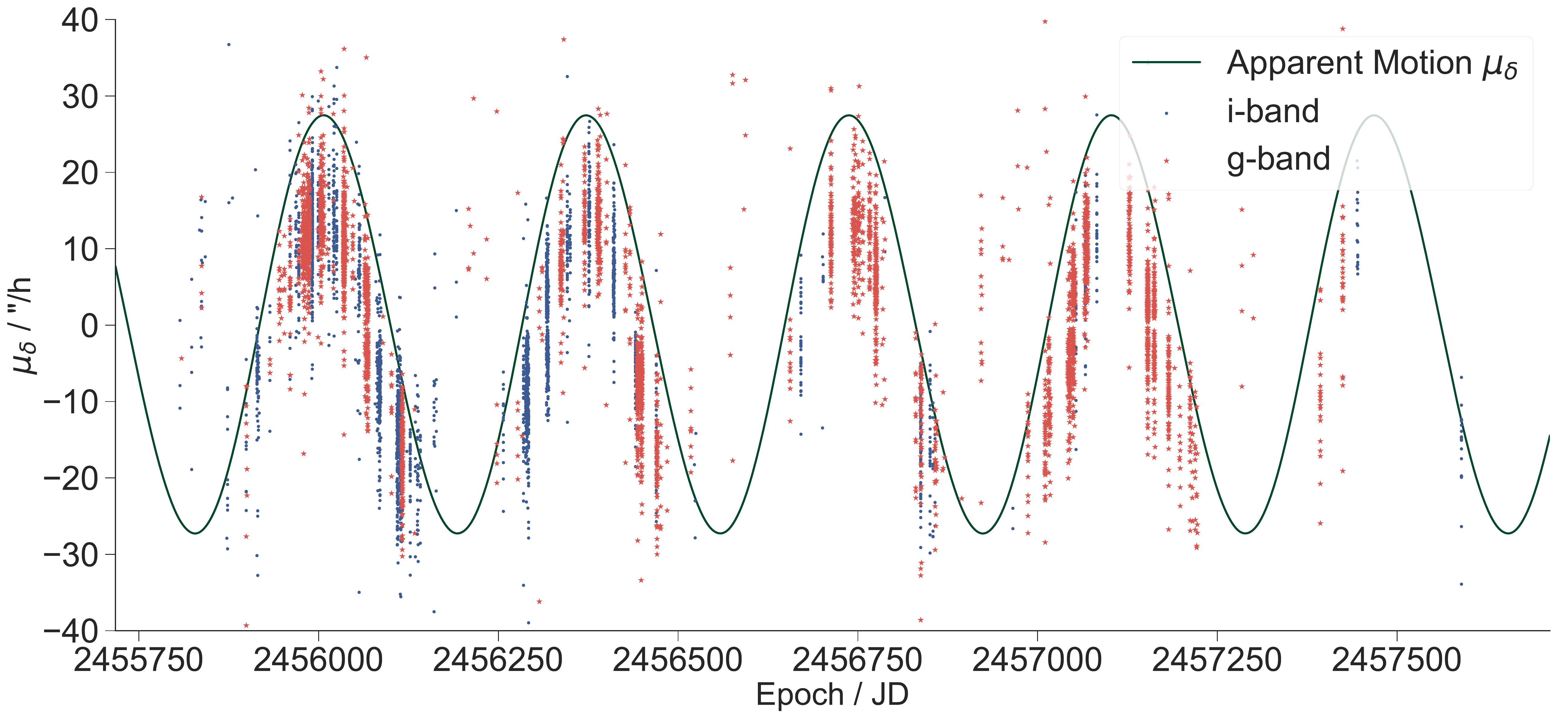}

    \caption{\inl{Observation epochs of KiDS in the $g$ and $i$ band vs. time:} The distributions of observation epochs for the different filters result in different distributions of the proper motion of SSOs (Fig.~\ref{pm_kde_ugri}). The red stars mark the observed proper motions in declination over time for all SSOs observed in the $g$ band. The blue dots show the same distribution for the $i$ band. The apparent motion component of this proper motion due to Earth's movement about the Sun is shown as a green line. For details on this approximation, refer to text.}
    \label{fig:giVsTime}
\end{figure}

\begin{figure}[t]
    \centering
    \includegraphics[width=\columnwidth]{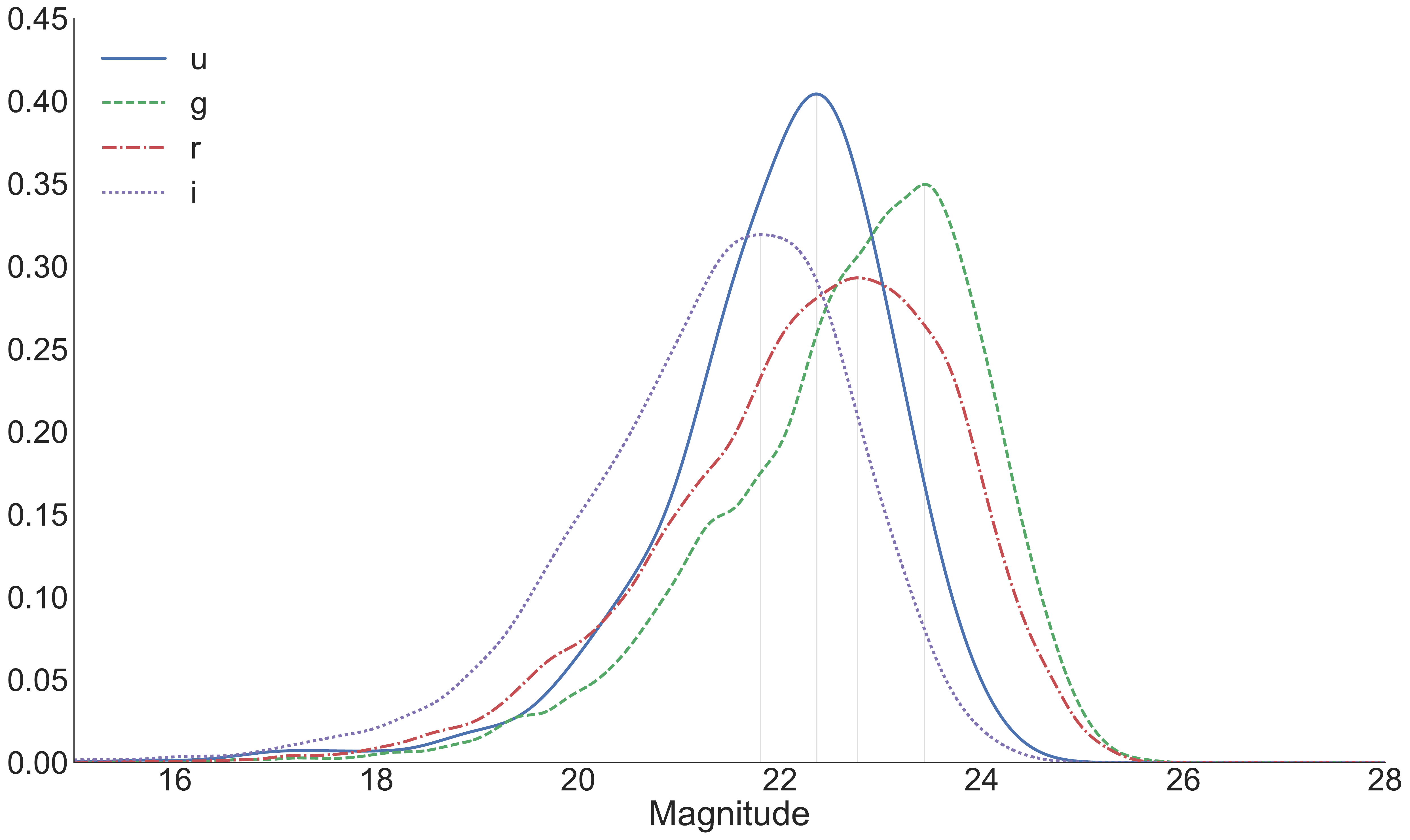}
    \caption{\inl{KDE of apparent Magnitude in different Filter}: The normalized distribution of the apparent magnitude of all recovered SSO candidates in the KiDS images in the different filters. Dashed lines indicate the derived limiting magnitude for SSO observation in each band, given by the peak of each distribution.}

    \label{mag_kde_ugri}
\end{figure}

Figure \ref{pm_kde_ugri} shows that the proper motion KDE behaves differently for all filters. We attribute this to the different distributions of Earth's orbital location across observations per filter. The apparent motion component in the proper motion in declination of SSOs varies with Earth's position, in both amplitude and sign. Figure~\ref{fig:giVsTime} shows this effect as an example for the observations in the $g$ and $i$ band. In the figure we show the observed proper motions in declination over time for all SSOs observed in the $g$ and $i$ bands. We also show the approximated apparent motion component of the measured proper motion, due to Earth's movement. Using IMCCE's Miriade VO tool, we computed the proper motion of the Sun in declination as seen from Earth, and adjusted the amplitude of this modulation to a distance of 1.7\,AU, the approximated distance of the main belt to the Earth. As the apparent motion in declination is caused by Earth's rotational axial tilt with respect to the ecliptic, we applied a 180$\,^{\circ}$ phase-shift, to account for the fact that we are observing toward the outer solar system, not inwards toward the Sun. The resulting modulation fits the observed dependency in the proper motion values well, even though we applied several approximations. We can see that the observation epochs introduce a periodic signal into the proper motion distributions of the SSOs. This modulation affects the proper motion distributions plotted in Fig.\,\ref{pm_kde_ugri}.

Figure\,\ref{mag_kde_ugri} shows the normalized distribution of magnitudes of all SSOs per band. The limiting magnitudes, derived by the peak position of the distribution for each band, correspond well to the limiting magnitudes ranges for the single exposures in each band, as given in Table\,\ref{filter}.

\subsection{Comparison with known SSOs}

We can list the distribution of the cross-matched KiDS SSOs over the dynamical classes of SSO. We differentiate between the following populations of SSOs:
\begin{itemize}
    \item main belt asteroids (MBA), the main reservoir of small bodies between Mars and Jupiter
    \item near-Earth objects (NEO), with orbits crossing that of terrestrial planets
    \item Mars-crossers (MC), a transitory population between the main-belt and the NEOs, which orbits intersect that of Mars
    \item Jupiter's Trojan asteroids (JTA), leading or preceding Jupiter on its orbit on the L4 and L5 Lagrangian points of the Sun-Jupiter system
    \item Centaurs (Cen), orbiting the Sun between the outer planets
    \item Kuiper-belt objects (KBO), with semi-major axes greater than that of Neptune
    \item Comets (Com), active bodies on highly eccentric orbits
\end{itemize}

The distribution is listed in Table~\ref{tab:class_distribution}. 97.7\% of the cross-matched SSOs in our sample are MBAs. Noteworthy is the detection of at least 4 KBOs, which are on the outskirts of the solar system and therefore faint and difficult to detect.

\begin{table}[t]
\centering
\caption{Distribution of KiDS SSOs over SSO populations: By cross-matching our sample with the SkyBoT database, we identified 10\,793 objects, making up 53.4\,\% of the sample. The distribution of the identified objects over the SSO populations as given in the text are listed here, stating the number of identified objects for each class and the fraction in relation to the whole sample of identified SSOs. A simple extrapolation to the entire sample would give a factor of two more for each population.}
\label{tab:class_distribution}
    \begin{tabular}{lllllllll} \toprule
                 & MBA   & JTA & MC  & NEO    & KBO  & Cen & Com \\ \midrule
  Number     & 10\,542 & 150 & 81   & 13     & 4    & 1       & 2 \\
  Fraction / $\%$   & 97.7& 1.4&0.8 & 0.1& 0.04 & 0.01 & 0.02 \\
\midrule
 \end{tabular}
\end{table}

\subsection{Unknown SSOs in the sample}
\begin{figure}[t]
    \centering
    \includegraphics[width=1\columnwidth]{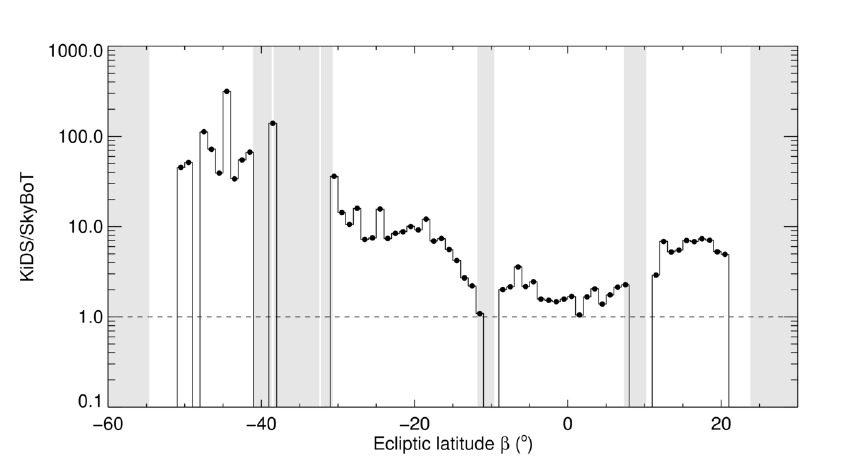}
    \caption{\inl{Ratio of SSOs per ecliptic latitude of KiDS SSOs and known SSOs:} The ratio of number of SSOs per ecliptic latitude per square degree for KiDS and SkyBot. The black dashed line marks the ratio equal to one. The observed areas in the sky span from $-51.6\,^{\circ}$ to $20.6\,^{\circ}$ ecliptic latitude. Gray shaded areas indicate ecliptic latitudes which were not covered by the survey. The KiDS sample has a much higher content of SSOs per square degree at high inclinations than the known SSOs in the SkyBot database.}
    \label{ecliptic_latitude}
\end{figure}

As shown in Fig.\,\ref{kids_layout}, the KiDS survey observes the sky both close to the ecliptic and in regions with high inclination. In Fig.\,\ref{ecliptic_latitude}, we compare the number of SSOs per square degree for the range of ecliptic latitudes covered in KiDS to the equivalent number of known SSOs in SkyBoT. In the figure we show the line at which the ratio of observed and predicted SSOs is equal to one. We can see that we recover the number of known SSOs in regions close to the ecliptic, with a smaller ratio toward the latitudes not covered in the KiDS survey (shown in the figure). Toward higher inclinations, we recover many more SSOs than previously known ones.

\begin{figure}[t]
    \centering
    \includegraphics[width=1\columnwidth]{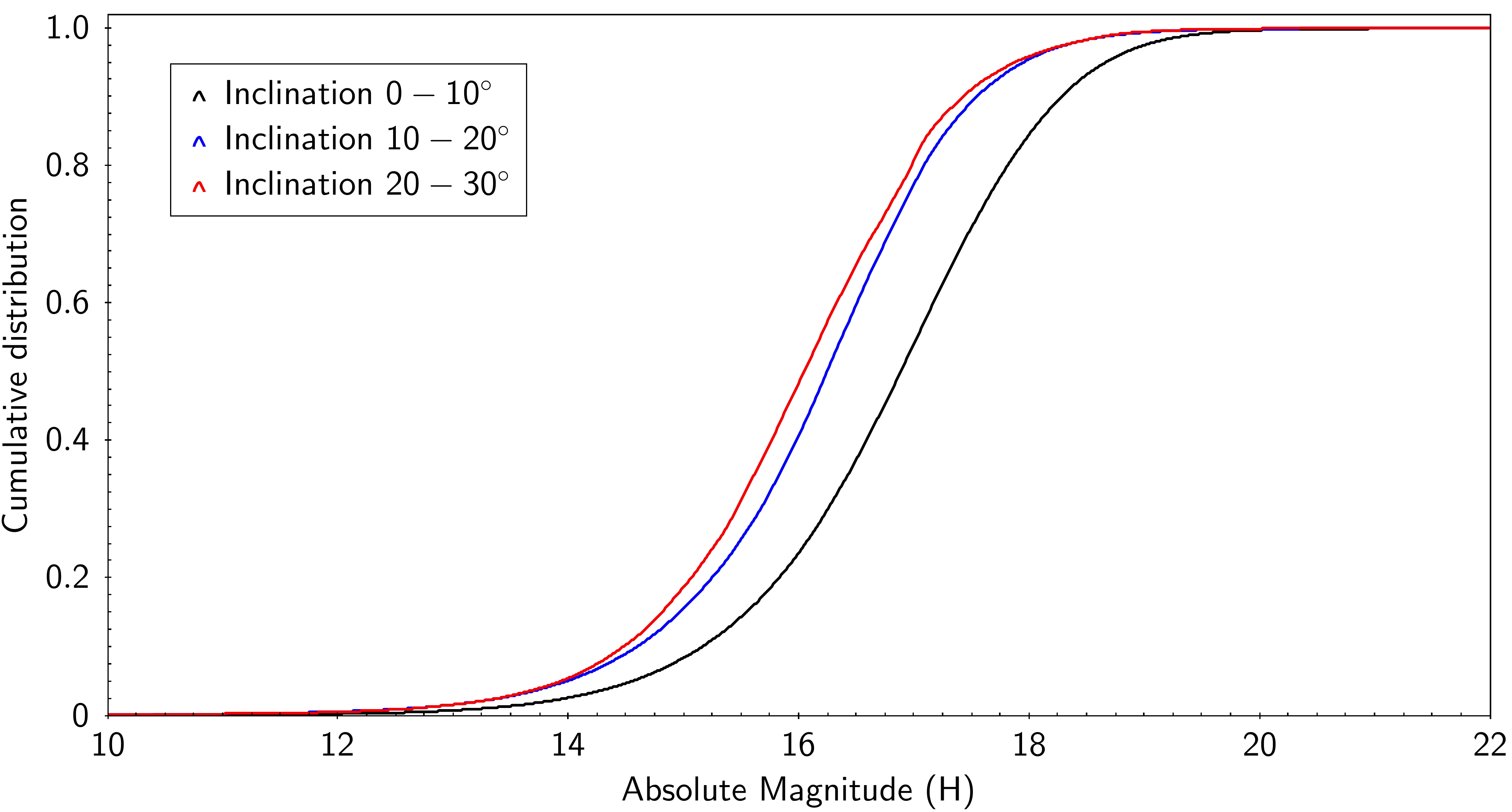}
    \caption{\inl{Discovery bias toward high inclinations:} The cumulative distributions of the absolute magnitudes of SSOs in the ASTORB database with orbit inclinations between 0$\,^{\circ}$-10$\,^{\circ}$ (black), between 10$\,^{\circ}$-20$\,^{\circ}$ (blue) and between 20$\,^{\circ}$-30$\,^{\circ}$ (red).}
    \label{sso_lat}
\end{figure}

 The search for SSOs is concentrated on regions close to the ecliptic, so the potential for discovery is larger when searching regions with high inclinations due to this bias.
This is shown in Fig.\,\ref{sso_lat}, where we compare the cumulative distributions of absolute magnitudes of SSOs in the Asteroid Orbital Elements Database\footnote{\url{ftp://ftp.lowell.edu/pub/elgb/astorb.html}} \citep[ASTORB,][]{1993-LPI-Bowell} for SSOs at different inclinations. A shift in the distribution toward brighter magnitudes for SSOs with higher inclinations is visible. This shows that the census for objects with high-inclinations is less complete, assuming a cumulative size distribution of the SSOs in the shape of a power-law \citep{2013AJ....146..111T, 1969JGR....74.2531D, gladman2009asteroid}.

\section{Conclusion and outlook}
We have presented a search for SSOs in 346\,${\rm deg}^2$ of the sky imaged by the Kilo-Degree Survey, corresponding to 65\,\% of the available data set and 23\,\% of the entire survey. We find 20\,221 SSO candidates, 53.4\,\% of which have a counterpart within 10\arcsec~in the SkyBot database.
Repeating the same analysis on the KiDS final release could therefore detect between 100,000 and 150,000 SSOs, half of them being potential discoveries.

The contamination of the complete sample is estimated to be lower than 0.05\,\%, while the degree of completeness is uncertain and would require further studies. By studying the distribution of artifacts, we found that most of them were observed only three times, and that they were located in close proximity to stars. Including all objects with three observations only and within 300\,\arcsec/\,h of stars increases the number of retrieved SSO candidates to 28\,290, while also increasing the false-positive content to 6.8\,$\pm$\,0.5\,\%.
The recovered SSOs are mostly main-belt objects, as expected. However, we also find near-Earth and trans-neptunian objects. Due to the KiDS survey design, we recover a large ratio of SSOs with high inclination.

The method can be easily ported to other large scale surveys, such as the other VST surveys and VISTA surveys, and would allow the discovery and recovery of large numbers of SSO. The present study was in particular motivated by ESA's future Euclid mission\footnote{\url{http://sci.esa.int/euclid/}} that will survey 15\,000\,${\rm deg}^2$ down to V=24.5, and is expected to observe several $10^5$ SSOs (Carry, submitted to A\&A).

Tracing an SSO detected in one field to other detections in neighboring fields or to the same field in a different filter could increase the amount of SSOs recovered (especially KBOs), give the rotation-induced light curves of the SSOs, and place higher constraints on the orbits of the objects. Due to the large cadence between the exposures in different filters of the square degree fields in KiDS, this linking of detections is a difficult task and beyond the scope of this work. In surveys containing observations that have a long sequence duration and small cadences between exposures of the same area of the sky, this could be a promising follow-up analysis.

\begin{acknowledgements}
Based on data products from observations made with ESO Telescopes at the La Silla Paranal Observatory under program IDs 177.A-3016, 177.A-3017 and 177.A-3018, and on data products produced by Target/OmegaCEN, INAF-OACN, INAF-OAPD and the KiDS production team, on behalf of the KiDS consortium. OmegaCEN and the KiDS production team acknowledge support by NOVA and NWO-M grants. Members of INAF-OAPD and INAF-OACN also acknowledge the support from the Department of Physics \& Astronomy of the University of Padova, and of the Department of Physics of Univ. Federico II (Naples).

This research made use of Astropy, a community-developed core Python package for Astronomy (Astropy Collaboration, 2013). This research has made use of the VizieR catalog access tool, CDS, Strasbourg, France.\\
We thank the anonymous referee for the valuable comments and suggestions.

Author Contributions: All authors contributed to the development and writing of this paper. The authorship list is given in two groups: the lead authors (MM, HB, BA, GVK, BC, EB), followed by an alphabetical group. The alphabetical group covers those who have either made a significant contribution to the data products, or to the scientific analysis.

GVK acknowledges financial support from the Netherlands Research School for Astronomy (NOVA) and Target. Target is supported by Samenwerkingsverband Noord Nederland, European fund for regional development, Dutch Ministry of economic affairs, Pieken in de Delta, Provinces of Groningen and Drenthe.

JTAdJ is supported by the Netherlands Organisation for Scientific Research (NWO) through grant 614.061.610.

KK acknowledges support by the Alexander von Humboldt Foundation.
\end{acknowledgements}
\bibliographystyle{aa}
\bibliography{bib}

\begin{appendix}
\section{SExtractor and SCAMP Configuration, SExtractor Parameter List}\label{app:scamp}
\subsection{SExtractor configuration file}

\begin{table}[H]
\begin{tabular}{ll}
\verb|DETECT_TYPE|      &   \verb|CCD|          \\
\verb|DETECT_MINAREA|   &   \verb|5|            \\
\verb|DETECT_MAXAREA|   &   \verb|0|            \\
\verb|THRESH_TYPE|      &   \verb|RELATIVE|     \\
                        &                           \\
\verb|DETECT_THRESH|    &   \verb|1.5|          \\
\verb|ANALYSIS_THRESH|  &   \verb|1.5|          \\
\verb|FILTER|           &   \verb|Y|           \\
\verb|FILTER_NAME|      &   \verb|gauss_2.5_5x5| \\
                        &                       \\
\verb|DEBLEND_NTHRESH|  &   \verb|16|          \\
\verb|DEBLEND_MINCONT|  &   \verb|0.05|        \\
\verb|CLEAN|            &   \verb|Y|           \\
\verb|CLEAN_PARAM|      &   \verb|1.0|         \\
\verb|MASK_TYPE|        &   \verb|CORRECT|      \\
\verb|WEIGHT_TYPE|      &   \verb|MAP_WEIGHT|   \\
\verb|RESCALE_WEIGHTS|  &   \verb|Y|               \\
\verb|WEIGHT_IMAGE|     &   \verb|Astro-WISE| \\
\verb|WEIGHT_GAIN|      &   \verb|Y|           \\ \midrule
\verb|PHOT_APERTURES|   &   \verb|25|          \\
\verb|PHOT_AUTOPARAMS|  &   \verb|1.0, 0.8|    \\
\verb|PHOT_PETROPARAMS| &   \verb|2.5, 3.5|    \\
\verb|PHOT_AUTOAPERS|   &   \verb|0.0,0.0|      \\
\verb|PHOT_FLUXFRAC|    &   \verb|0.5|          \\\midrule
\verb|SATUR_LEVEL|      &   \verb|60000.0|         \\
\verb|SATUR_KEY|        &   \verb|SATURATE|    \\
\verb|MAG_ZEROPOINT|    &   \verb|0.0|              \\
\verb|MAG_GAMMA|        &   \verb|4.0|              \\
\verb|GAIN|             &   \verb|2.5|          \\
\verb|GAIN_KEY|         &   \verb|GAIN|         \\
\verb|PIXEL_SCALE|      &   \verb|0.0|          \\ \midrule
\verb|SEEING_FWHM|      &   \verb|1.2|          \\
\verb|STARNNW_NAME|     &   \verb|default.nnw|  \\
\end{tabular}
\end{table}
\begin{table}[H]
\begin{tabular}{ll}
\verb|BACK_TYPE|        &   \verb|AUTO|             \\
\verb|BACK_VALUE|       &   \verb|0.0|          \\
\verb|BACK_SIZE|        &   \verb|64|           \\
\verb|BACK_FILTERSIZE|  &   \verb|3|                \\
\verb|BACKPHOTO_TYPE|   &   \verb|LOCAL|            \\
\verb|BACKPHOTO_THICK|  &   \verb|24|           \\
\verb|BACK_FILTTHRESH|  &   \verb|0.0|          \\
\end{tabular}
\end{table}

\subsection{SExtractor parameter list}
\begin{table}[H]
\begin{tabular}{lll}
\verb|NUMBER| & \verb|X_IMAGE| & \verb|Y_IMAGE| \\
\verb|X2_IMAGE| & \verb|Y2_IMAGE| & \verb|XY_IMAGE| \\
\verb|ISOAREA_IMAGE| & \verb|BACKGROUND| & \verb|FLAGS| \\
\verb|THRESHOLD| & \verb|FLUX_MAX| & \verb|A_IMAGE| \\
\verb|B_IMAGE| & \verb|THETA_IMAGE| & \verb|ERRA_IMAGE| \\
\verb|FLUX_ISO| & \verb|FLUXERR_ISO| & \verb|MAG_ISO| \\
\verb|MAGERR_ISO| & \verb|FLUX_APER| & \verb|FLUXERR_APER| \\
\verb|MAG_APER| & \verb|MAGERR_APER| & \verb|ALPHA_SKY| \\
\verb|DELTA_SKY| & \verb|ERRB_IMAGE| & \verb|ERRTHETA_IMAGE| \\
\verb|MU_MAX| & \verb|FWHM_IMAGE| & \verb|CLASS_STAR| \\
\verb|FLUX_RADIUS| & \verb|ELONGATION| & \verb|ELLIPTICITY| \\
\verb|CXX_IMAGE| & \verb|CXY_IMAGE| & \verb|CYY_IMAGE| \\
\verb|ERRCXX_IMAGE| & \verb|ERRCXY_IMAGE| & \verb|ERRCYY_IMAGE| \\
\verb|MAG_AUTO| & \verb|XWIN_IMAGE| & \verb|YWIN_IMAGE| \\
\verb|FLUX_AUTO| & \verb|FLUXERR_AUTO| & \verb|MAGERR_AUTO| \\
\verb|ALPHA_J2000| & \verb|DELTA_J2000| & \verb|ERRX2_WORLD| \\
\verb|ERRY2_WORLD| & \verb|ERRXY_WORLD| & \verb|AWIN_IMAGE| \\
\verb|BWIN_IMAGE| & \verb|THETAWIN_IMAGE| & \verb|ERRAWIN_IMAGE| \\
\verb|ERRBWIN_IMAGE| & \verb|ERRTHETAWIN_IMAGE| & \verb|FWHM_WORLD|
\end{tabular}
\end{table}

\subsection{SCAMP configuration file}
\begin{table}[H]
\begin{tabular}{ll}
\verb|FGROUP_RADIUS|      &   \verb|1.0|    \\\midrule
\verb|REF_SERVER|      &   \verb|cocat1.u-strasbg.fr|\\
\verb|ASTREF_CATALOG|      &   \verb|2MASS|         \\\midrule
\verb|ASTREF_BAND|      &   \verb|DEFAULT|       \\
\verb|ASTREFCENT_KEYS|      &   \verb|X_WORLD,Y_WORLD|\\
\verb|ASTREFERR_KEYS|      &   \verb|ERRA_WORLD, ERRB_WORLD| \\
                            &   \verb|ERRTHETA_WORLD|\\
\verb|ASTREFMAG_KEY|      &   \verb|MAG|           \\
\verb|ASTREFMAGERR_KEY|      &   \verb|MAGERR|        \\
\verb|ASTREFOBSDATE_KEY|      &   \verb|OBSDATE|       \\
\verb|ASTREFMAG_LIMITS|      &   \verb|-99.0,99.0|    \\
\verb|MATCH|      &   \verb|Y|               \\
\verb|MATCH_NMAX|      &   \verb|0|               \\
\verb|PIXSCALE_MAXERR|      &   \verb|1.0|             \\
\verb|POSANGLE_MAXERR|      &   \verb|2.0|             \\
\verb|POSITION_MAXERR|      &   \verb|0.017|           \\
\verb|MATCH_RESOL|      &   \verb|0|               \\
\verb|MATCH_FLIPPED|      &   \verb|N|               \\
\verb|MOSAIC_TYPE|      &   \verb|UNCHANGED|       \\
\verb|FIXFOCALPLANE_NMIN|      &   \verb|3|             \\
\end{tabular}
\end{table}

\begin{table}
\begin{tabular}{ll}

\verb|CROSSID_RADIUS|      &   \verb|10.0|           \\\midrule
\verb|SOLVE_ASTROM|      &   \verb|Y|               \\
\verb|PROJECTION_TYPE|      &   \verb|SAME|            \\
\verb|ASTRINSTRU_KEY|      &   \verb|FILT_ID,CHIP_ID|  \\
\verb|STABILITY_TYPE|      &   \verb|INSTRUMENT|      \\
\verb|CENTROID_KEYS|      &   \verb|XWIN_IMAGE,YWIN_IMAGE| \\
\verb|CENTROIDERR_KEYS|      &   \verb|ERRAWIN_IMAGE,ERRBWIN_IMAGE,| \\
                            &\verb|ERRTHETAWIN_IMAGE|\\
\verb|DISTORT_KEYS|      &   \verb|XWIN_IMAGE,YWIN_IMAGE| \\
\verb|DISTORT_GROUPS|      &   \verb|1,1|             \\
\verb|DISTORT_DEGREES|      &   \verb|2|               \\
\verb|FOCDISTORT_DEGREE|      &   \verb|1|               \\
\verb|ASTREF_WEIGHT|      &   \verb|1.0|             \\ \midrule
\verb|ASTRACCURACY_TYPE|      &   \verb|TURBULENCE-ARCSEC| \\
\verb|ASTRACCURACY_KEY|      &   \verb|ASTRACCU|        \\
\verb|ASTR_ACCURACY|      &   \verb|0.054|            \\
\verb|ASTRCLIP_NSIGMA|      &   \verb|1.5|             \\
\verb|COMPUTE_PARALLAXES|      &   \verb|N|               \\
\verb|COMPUTE_PROPERMOTIONS|   &   \verb|Y|               \\
\verb|CORRECT_COLOURSHIFTS|    &   \verb|N|               \\
\verb|INCLUDE_ASTREFCATALOG|   &   \verb|Y|               \\
\verb|ASTR_FLAGSMASK|          &   \verb|0x00fc|          \\
\verb|ASTR_IMAFLAGSMASK|       &   \verb|0x0|             \\\midrule
\verb|SOLVE_PHOTOM|            &   \verb|Y|               \\
\verb|MAGZERO_OUT|             &   \verb|0.0|             \\
\verb|MAGZERO_INTERR|          &   \verb|0.01|            \\
\verb|MAGZERO_REFERR|          &   \verb|0.03|            \\
\verb|PHOTINSTRU_KEY|          &   \verb|FILT_ID|         \\
\verb|MAGZERO_KEY|             &   \verb|ZEROPNT|          \\
\verb|EXPOTIME_KEY|            &   \verb|EXPTIME|         \\
\verb|AIRMASS_KEY|             &   \verb|AIRMASS|         \\
\verb|EXTINCT_KEY|             &   \verb|PHOT_K|          \\
\verb|PHOTOMFLAG_KEY|          &   \verb|PHOTFLAG|        \\
\verb|PHOTFLUX_KEY|            &   \verb|FLUX_AUTO|       \\
\verb|PHOTFLUXERR_KEY|         &   \verb|FLUXERR_AUTO|    \\
\verb|PHOTCLIP_NSIGMA|         &   \verb|1.5|             \\
\verb|PHOT_ACCURACY|           &   \verb|1e-3|            \\
\verb|PHOT_FLAGSMASK|          &   \verb|0x00fc|          \\
\verb|PHOT_IMAFLAGSMASK|       &   \verb|0x0|             \\\midrule
\verb|SN_THRESHOLDS|           &   \verb|3.0,100.0|      \\
\verb|FWHM_THRESHOLDS|         &   \verb|0.0,1000.0|       \\
\verb|ELLIPTICITY_MAX|         &   \verb|1.0|             \\
\verb|FLAGS_MASK|              &   \verb|239|          \\

\verb|WEIGHTFLAGS_MASK|        &   \verb|0x00ff|          \\
\verb|IMAFLAGS_MASK|           &   \verb|0x0|            \\

\end{tabular}
\end{table}

\end{appendix}

\end{document}